\documentclass[aps,pra,twocolumn]{revtex4}
\usepackage[dvips]{epsfig}
\usepackage[usenames]{color}
\usepackage{pstricks}
\usepackage{amsmath}
\usepackage{amssymb}
\usepackage{subfigure}
\usepackage{afterpage}
\usepackage{multirow}

\begin{document}

\title{Ultracold molecular collisions in combined electric and magnetic fields}

\author{Goulven Qu{\'e}m{\'e}ner}
\affiliation{Laboratoire Aim\'e Cotton, CNRS, Universit\'e Paris-Sud, ENS
Cachan, 
Campus d'Orsay, B\^atiment 505, 91405 Orsay, FRANCE} 
\email{goulven.quemener@u-psud.fr}
\author{John L. Bohn}
\affiliation{JILA, NIST and University of Colorado,
Boulder, C0 80309, USA} 

\date{\today}

\begin{abstract}

We consider collisions of electric and magnetic polar molecules, 
taking the OH radical as an example, subject to combined electric and 
magnetic static fields. We show that the relative orientation of the 
fields has an important effect on the collision processes for different 
fields magnitude at different collision energies. 
This is due to the way the molecules polarize in 
the combined electric and magnetic fields and hence 
the way the electric dipole-dipole interaction rises.
If OH molecules are confined in magnetic quadrupole traps
and if an electric field is applied, molecular collisions will strongly
depend on the position as well as the velocity of the molecules,
and consequences on the molecular dynamics are discussed.

\end{abstract}


\maketitle

\font\smallfont=cmr7

\section{Introduction}

Cold molecules, with translational temperatures at or below 100 mK, are strongly
subject to control over their behavior, and may afford unprecedented
opportunities for probing chemistry as a function of initial conditions to
reaction~\cite{Quemener_CR_112_4949_2012}.  
Thus, for example, $\mu$K samples of KRb molecules have been
formed~\cite{Ni_S_322_231_2008} and their reactions probed for 
different temperatures~\cite{Ospelkaus_S_327_853_2010}, electric
fields~\cite{Ni_N_464_1324_2010} and dimensional
confinements~\cite{DeMiranda_NP_7_502_2011,Chotia_PRL_108_080405_2012}.  
These molecules have appreciable electric dipole moments, so manipulation
of their collisions arises from their comparatively strong dipolar
interactions.

More broadly, open-shell radicals can also be produced at low temperatures, 
albeit in samples not quite as cold.  Examples include $^2\Sigma$ molecules 
such as SrF~\cite{Shuman_N_467_820_2010} or $^2\Pi$ molecules such 
as OH ~\cite{Stuhl_N_492_396_2012}. 
In addition to being of arguably greater
chemical interest, these species present the possibility of simultaneous control
by acting on their magnetic, as well as electric, dipole moments.  
The simultaneous action of electric and magnetic fields has been considered
previously in the context of buffer-gas-cooled species, 
considering collisions such as He + CaD and He +
ND~\cite{Tscherbul_JCP_125_194311_2006,Abrahamsson_JCP_127_044302_2007,
Tscherbul_JCP_128_244305_2008} or He +
YbF~\cite{Tscherbul_PRA_75_033416_2007}. For certain radicals, such as OH, O$_2$
and NH, molecule-molecule collisions have been considered in the presence of
either 
electic~\cite{Avdeenkov_PRA_66_052718_2002} or
magnetic~\cite{Ticknor_PRA_71_022709_2005,Tscherbul_NJP_11_055021_2009,
Perez-Rios_JCP_134_124310_2011,Janssen_PRA_83_022713_2011,
Suleimanov_JCP_137_024103_2012,Janssen_PRL_110_063201_2013} 
fields but not, to our knowledge, both simultaneously.

Here we consider the effect of both electric and magnetic  fields on collisions
of the OH radical, at collision energies ranging from 1$\mu$K to 50
mK. 
In its ground electronic state, this molecule possesses a magnetic 
dipole moment of $|\vec{\mu}| = 2 \, \mu_B $ ($\mu_B$ is the Bohr magneton) 
and an electric dipole moment 
of $|\vec{d}| = 1.67$~D.   Thus at the temperatures considered, long-range
electric
dipole forces 
generate interaction energies that can exceed translational temperatures when
the molecules are hundreds of Bohr radii apart.  This circumstance implies that 
electric and magnetic fields act on the molecules primarily on this distance
scale, and that theoretical models focusing on this long-range physics are
adequate to see the effect of the fields.  From this standpoint, it has already
been noted that electric fields tend to {\it increase} the rate of
state-changing collisions of OH molecules~\cite{Avdeenkov_PRA_66_052718_2002},
while magnetic fields tend to {\it decrease} these
rates~\cite{Ticknor_PRA_71_022709_2005}. If both types of field are present,
they are therefore in competition, promising additional opportunities for
manipulation of collisions.  In particular, the angle between the fields, at the
site of a collision, can be decisive in determining the collision's outcome.

The emphasis on long-range physics is assisted by the special 
characteristics of $^2\Pi$ molecules such as OH.  For $\Sigma$ molecules, 
the electric dipole moment is induced by the mixing of the ground and 
the excited rotational states by an electric field. The rotational 
constant is on the order of mK so the ground and higher excited rotational 
levels can not be treated independently, while for OH molecules the large 
rotation splitting implies small mixing of higher-lying rotational states
at modest electric fields. 
A signature of this feature is the protection of certain low field seeking 
states of cold OH molecules in a magnetic field, leading to high elastic 
collisions compared to inelastic collisions, stemming from a strong repulsive 
van der Waals coefficient~\cite{Stuhl_N_492_396_2012}.  Because of this
repulsion, 
the OH molecules in those states are expected to be shielded from chemical
reactions at sufficiently low temperature.

In this paper, we investigate the scattering of polar molecules when 
arbitrary combined electric and magnetic fields are applied (parallel 
as well as non-parallel fields), taking the OH molecule as an example.
This study is the starting point to more complicated dynamics of polar 
molecules in a quadrupole magnetic trap in a presence of an electric 
field as performed in ongoing experiments~\cite{Stuhl_MP_2013}, where 
collisions of molecules will occur at different electric and magnetic 
field configurations for different positions in the trap.
These collisions are also important to determine new efficient evaporative cooling 
schemes, for example using appropriate electric and magnetic field trap 
combinations, to further cool down molecular dipolar gases. 
If quantum degenerate gases are finally produced, the combined electric 
and magnetic fields can be used as additional tools to control and probe 
the many-body physics of electric and magnetic dipolar 
systems~\cite{Gorshkov_PRL_107_115301_2011,Baranov_CR_112_5012_2012,Wall_arXiv_1212_3042_2012}.

The paper is organized as follows. In Section II, we describe the time-independent 
quantum formalism 
used to perform the scattering calculations, presented in section III. 
We conclude in Section IV.  

\section{Scattering in combined electric and magnetic fields}

We present here the time-independent quantum formalism used in 
this work for the scattering of two OH molecules in arbitrary 
combined electric fields. 

\subsection{Molecular energies and functions}

The OH molecule in its ground rovibronic state $^2\Pi_{3/2}, v=0, j=3/2$ 
is well described by a Hund's case (a) scheme. $j$ is the quantum number 
associated with its rotational angular momentum $\vec{j}$,  $m_j$ is its projection onto the laboratory space-fixed 
axis with unit vector $\hat{Z}$, and $\omega_j$ is its projection onto the
molecular body-fixed axis with unit vector $\hat{z}$. 
In its ground state, $\omega_j=\pm3/2$ is the sum of $\lambda=\pm1$ and $\sigma=\pm1/2$, the
values of the projection of the electronic orbital $\vec{l}$ and spin $\vec{s}$ angular momentum
onto the body-fixed axis.
A good basis set for Hund's case (a) molecule is therefore
$|j, m_j, \omega_j \rangle |\lambda, \sigma \rangle$~\cite{Brown_Carrington_Book_2003}.
The molecule exhibits a small Lambda-doublet of $\Delta\approx 80$~mK between
two states $e$ and $f$ of different parity within its ground rovibronic 
state $^2\Pi_{3/2}, v=0, j=3/2$. The electric field mixes these two 
states to induce the electric dipole moment in the laboratory frame. 
The next rotational level $^2\Pi_{3/2}, v=0, j=5/2$ is $\simeq 100$~K
higher than the $j=3/2$
state so the OH ground rotational state is well-separated from all 
its higher excited states, and will be ignored in the rest of the study 
considering the low collision energy range of the molecules.

\begin{figure} [b]
\begin{center}
\includegraphics*[width=6cm,keepaspectratio=true,angle=-90]{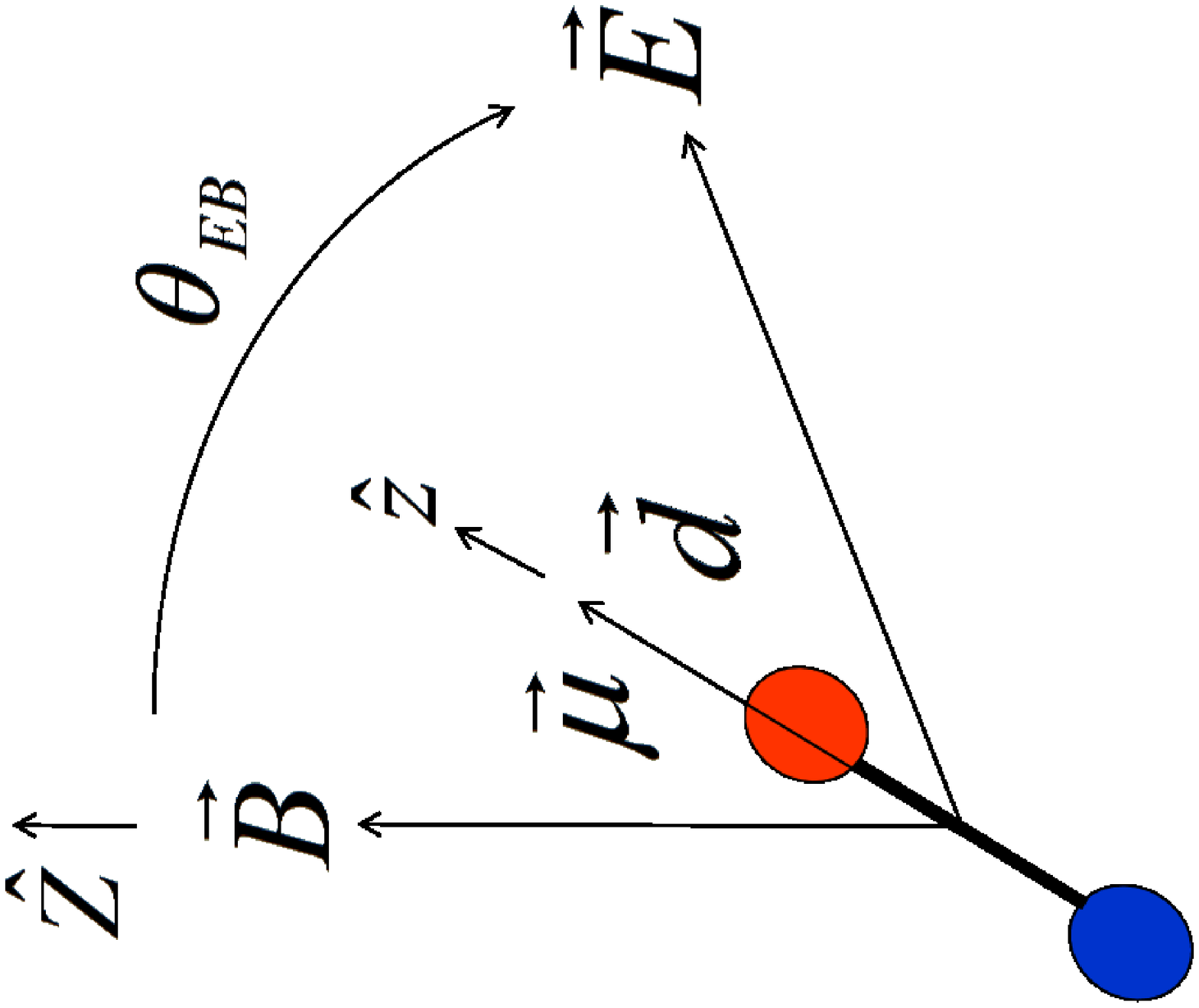}
\caption{(Color online) Schematic electric and magnetic dipole moments 
for an OH molecule in a Hund's case (a) scheme, in the presence of an 
arbitrary electric and magnetic field configuration $\vec{E}, \vec{B}, \theta_{EB}$.
\label{MOLECULE-FIG}
}
\end{center}
\end{figure}

In Hund's case (a), the magnetic dipole moment is given to a good approximation by 
$\vec{\mu} = - \mu_B \, (g_{s} \sigma + g_{l} \lambda) \, \hat{z}$
where $\mu_B$ is the Bohr magneton, $g_s$ is the electron's g factor 
($g_s \sim 2.002 \approx 2$) and $g_{l}=1$, so that 
$ \vec{\mu} \approx \pm 2 \, \mu_B \, \hat{z}$.
The electric dipole moment is given by $\vec{d} = d \, \hat{z}$ with $d = 1.67$~D.
As a consequence, in a Hund's case (a) scheme, both electric and 
magnetic dipole moments lie along the molecular axis, as depicted 
schematically on Fig.~\ref{MOLECULE-FIG} (if the two dipoles point in the same direction).
This implies that we do not take into account couplings 
between the states $|\omega_j|=3/2$ and $|\omega_j|=1/2$. This is not important 
in this study since the first $|\omega_j|=1/2$ state lies well above the ground 
state by more than $100$~K~\cite{Brown_Carrington_Book_2003}.

When a magnetic field is applied, the interaction of the molecule with 
the field is given by the Zeeman Hamiltonian
$H_{\rm{Z}} = -\vec{\mu} \cdot \vec{{B}} $.
%
%
In the Hund's case (a) basis set $|j, m_j, \omega_j \rangle$ (we ignore the 
spectator ket $|\lambda, \sigma \rangle$ in the following unless stated
otherwise and we take $\omega_j=\omega_j'$), 
it takes the form
%
%
\begin{multline}
\langle j, m_j, \omega_j |
 \,  H_{\rm{Z}}\,  
| j', m_j', \omega_j \rangle 
 = - \mu \, {B}  \\
(-1)^{m_j-\omega_j} \,
\sqrt{2j+1} \, \sqrt{2j'+1}  \\ 
\left( \begin{array}{ccc} j & 1 & j' \\ -\omega_j & 0 & \omega_j \end{array}
\right)
\left( \begin{array}{ccc} j & 1 & j' \\ -m_j & 0 & m_j' \end{array} 
\right) 
\label{Zeeman}
\end{multline}
with $\mu = - \mu_B \, (g_{s} \sigma + g_{l} \lambda)$.
Here we choose $\vec{{B}}$ to point along the space-fixed frame
axis $\hat{Z}$ (as shown in Fig.~\ref{MOLECULE-FIG}).
Based on the symmetry of the three-$j$ symbols, we must have $j'-j=\Delta j=0,\pm1$, 
$m_j' = m_j$.

When an electric field is applied, the interaction of the molecule with 
the field is given by the Stark Hamiltonian
$H_{\rm{S}} = -\vec{d} \cdot \vec{{E}} $
%
%
and in the Hund's case (a) basis set, it takes the form
%
%
\begin{multline}
\langle j, m_j, \omega_j |
 \, H_{\rm{S}} \,  
| j', m_j', \omega_j \rangle \\
 = - d \, {E} \,
\sqrt{\frac{4\pi}{3}} Y_{1,m_j'-m_j}(\theta_{EB},\phi_{EB}) \\
(-1)^{m_j-\omega_j} \, 
\sqrt{2j+1} \, \sqrt{2j'+1} \\
\left( \begin{array}{ccc} j & 1 & j' \\ -\omega_j & 0 & \omega_j \end{array}
\right)
\left( \begin{array}{ccc} j & 1 & j' \\ -m_j & m_j - m_j' & m_j' \end{array}
\right)
\label{Stark}
\end{multline}
if we allow $\vec{{E}}$ to point in an arbitrary direction 
$(\theta_{EB},\phi_{EB})$ from the $\hat{Z}$ axis (as shown in Fig.~\ref{MOLECULE-FIG}). 
In the following we will set $\phi_{EB}=0$.
From the three-$j$ symbols, we must have again $j'-j=\Delta j=0,\pm1$. 
But now, if there is a nonzero  angle $\theta_{EB}$
between the $\vec{{E}}$ field and the $\vec{{B}}$ field, we have in 
general $m_j' - m_j = 0, \pm1$.
Thus the quantum numbers $m_j$ referred to a particular axis 
(the $\vec{{B}}$ or the $\vec{{E}}$ axis) are no longer
good. Good quantum numbers can be found along two particular
axes as shown in Ref.~\cite{Bohn_MP_2013}, but we do not do so
here. Note that if $\theta_{EB}=\pi/2$, $m_j' = m_j \pm1$ and 
if $\theta_{EB}=0$, we recover the case $m_j' = m_j$.

In the absence of rotation, molecules with the quantum numbers $\lambda=\pm1$ have the same energy 
causing a Lambda-doubling degeneracy for $\lambda \ge 1$. An additional term 
$H_{\lambda}$ stemming from the coupling of the rotation and the electronic angular 
momentum of the molecule splits this Lambda-doubling into two distinct states e and f 
of different parity~\cite{Brown_Carrington_Book_2003}. A good basis set is then 
the parity basis set 
%
%
\begin{multline}
|j, m_j, |\omega_j|, \epsilon \rangle \, | |\lambda|, |\sigma| \rangle = 
\dfrac{1}{\sqrt{2}} \, \bigg\{ 
|j, m_j, \omega_j \rangle \, | \lambda, \sigma \rangle \\
+ \epsilon \, |j, m_j, -\omega_j \rangle \, | -\lambda, -\sigma \rangle 
\bigg\}
\label{MSsym}
\end{multline}
with $\epsilon=\pm1$ corresponding respectivelly to the e/f states
parities~\cite{Lara_PRA_78_033433_2008}.
For OH, the f/e splitting is about $\Delta \approx 80$~mK and is diagonal in 
the parity basis,
%
%
\begin{multline}
\langle j, m_j, |\omega_j|, \epsilon |
 \, H_{\Lambda} \,  
| j', m_j', |\omega_j|, \epsilon' \rangle 
 = \\
 (-\epsilon) \, \Delta/2 \  \delta_{j, j'} \,  \delta_{m_j, m_j'} 
\, \delta_{\epsilon,\epsilon'}.
\label{Lambdaparity}
\end{multline}
The Zeeman expression~\eqref{Zeeman} in this new basis set is given by
%
%
\begin{multline}
\langle j, m_j, |\omega_j|, \epsilon |
 \,  H_{\rm{Z}}\,   
| j', m_j', |\omega_j|, \epsilon' \rangle 
 = 
 - \mu \, {B} \ \delta_{\epsilon, \epsilon'} \\
(-1)^{m_j-|\omega_j|} \,
\sqrt{2j+1} \, \sqrt{2j'+1} \\
\left( \begin{array}{ccc} j & 1 & j' \\ -|\omega_j| & 0 & |\omega_j|
\end{array}
\right)
\left( \begin{array}{ccc} j & 1 & j' \\ -m_j & 0 & m_j' \end{array}
\right) 
\label{Zeemanparity}
\end{multline}
with $\mu = - \mu_B \, (g_{s} |\sigma| + g_{l} |\lambda|) \approx - 2 \, \mu_B $.
The magnetic field mixes
states of the same parity $\epsilon=\epsilon'$ only.
The Stark expression~\eqref{Stark} is given by
%
%
\begin{multline}
\langle j, m_j, |\omega_j|, \epsilon |
 \,  H_{\rm{S}}\,   
| j', m_j', |\omega_j|, \epsilon' \rangle 
 = 
 - d \, {E} \, \ (1-\delta_{\epsilon,\epsilon'}) \\ 
\sqrt{\frac{4\pi}{3}} Y_{1,m_j'-m_j}(\theta_{EB},\phi_{EB}) \\
(-1)^{m_j-\omega_j} \, 
\sqrt{2j+1} \, \sqrt{2j'+1} \\
\left( \begin{array}{ccc} j & 1 & j' \\ -|\omega_j| & 0 & |\omega_j|
\end{array}
\right)
\left( \begin{array}{ccc} j & 1 & j' \\ -m_j & m_j - m_j' & m_j' \end{array}
\right)
\label{Starkparity}
\end{multline}
%
The electric field mixes
states of the different parity $\epsilon \ne \epsilon'$ only.
The selection rules $\Delta j=0,\pm1$ and $\Delta m_j=0,\pm1$ still hold.
In the following, we take $j=j'=3/2$.

\begin{figure} [t]
\begin{center}
\includegraphics*[width=6cm,keepaspectratio=true,angle=-90]{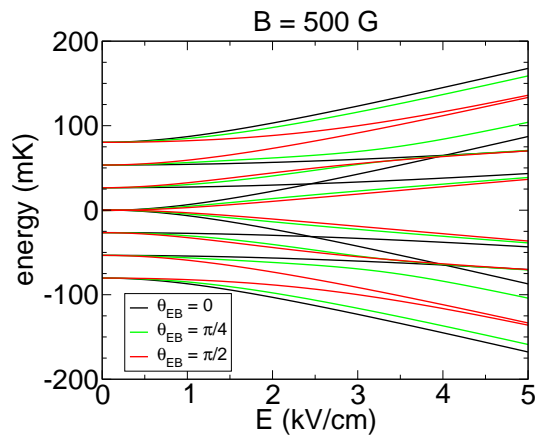}
\caption{(Color online) Eigenenergies of the ground state OH molecule
in combined electric and magnetic fields as a function of the electric field for
$\theta_{EB}=0$ (black curve), 
$\theta_{EB}=\pi/4$ (green curve), $\theta_{EB}=\pi/2$ (red curve). 
The magnetic field is $B=500$~G.
\label{NRG-FIG}
}
\end{center}
\end{figure}

The diagonalization of the molecular Hamiltonian $H_\lambda + H_{\rm{Z}} +
H_{\rm{S}}$ 
in the parity basis set given by the 
expressions~\eqref{Lambdaparity},~\eqref{Zeemanparity}, 
and~\eqref{Starkparity} leads to the eight eigenenergies 
denoted $\varepsilon_i$ with $i=1,...,8$ from lowest to highest energy; and
eigenfunctions 
denoted $|i\rangle$ of the OH molecule in a combined electric 
and magnetic field with a relative orientation $\theta_{EB}$. 
These energies are shown in Fig.~\ref{NRG-FIG} for a fixed magnetic 
field of $B = 500$~G as a function of the electric field $E$ for 
the three angles $\theta_{EB}=0$ (black curves, parallel fields), 
$\theta_{EB}=\pi/4$ (green curves, neither parallel, nor perpendicular), 
and $\theta_{EB}=\pi/2$ (red curves, perpendicular fields).

For the highest excited adiabatic state $|8\rangle$ represented by the 
highest energy curve, 
it is easier to induce 
an electric dipole moment when the electric field is more parallel to the 
magnetic field axis while it is harder when the fields are more perpendicular. 
This can be seen from the derivative of the energy curves with respect to 
the electric field which is proportional to the induced electric dipole moment. 
The derivative for the $\theta_{EB}=\pi/2$ red curve is smaller than that for 
the $\theta_{EB}=0$ black curve, the one for the $\theta_{EB}=\pi/4$ green 
curve sitting in between.
In terms of the results of Ref.~\cite{Bohn_MP_2013}, the induced electric
dipole moment $\tilde{d}$ for the upper adiabatic state $|8\rangle$ can be approximated by
\begin{eqnarray}
\tilde{d} \approx \frac{ | \omega_j m_j | }{ j(j+1) }
\left( 1 + \frac{ \mu B }{ d E } \cos( \theta_{EB} )\right) \, d.
\end{eqnarray}
Then, if $\theta_{EB}$ goes from 0 to $\pi/2$, $\tilde{d}$ decreases in magnitude 
for fixed $B$ and $E$ fields as seen on Fig.~\ref{NRG-FIG}.

\subsection{Molecular scattering}

In what follows we will be concerned with molecules colliding in their stretched
states $|8 \rangle$, which are magnetically trapped.  
In collisions, these molecules will exert torques on one another that can
disturb their orientation, producing molecules in states $|i<8 \rangle$ and in
general leading to trap loss and heating.  As these appear to be the dominant
loss collisions~\cite{Stuhl_N_492_396_2012}, we focus on them and ignore the
possibility of chemical reactions.  Our scattering theory is therefore similar
to the long-range-dominated theories in
Refs.~\cite{Avdeenkov_PRA_64_052703_2001,Tscherbul_NJP_11_055021_2009}.

We consider two OH molecules of mass $m_1, m_2$ and position $\vec{r}_1, \vec{r}_2$
respectively. We decouple the motion of the two-body system into a motion 
of a center of mass, of total mass $m_\text{tot} = m_1 + m_2$ and
position 
$\vec{R} = (m_1 \vec{r}_1 + m_2 \vec{r}_2) / (m_1+m_2)$, and a motion of a 
relative particle, of reduced mass $m_\text{red} = (m_1 m_2) / (m_1 + m_2)$ 
and position $\vec{r} = \vec{r}_2 - \vec{r}_1$.
The total Hamiltonian of the relative motion is $H = T + V$
with $T$ being the relative kinetic energy operator
of the relative motion and $V$ the potential energy.
The long-range electric dipole-dipole interaction is given by
\begin{eqnarray}
 V = \frac{{\vec{d}_1} \cdot {\vec{d}_2} 
- 3 ({\vec{d}_1} \cdot \hat{r}) ({\vec{d}_2} \cdot \hat{r})}{ 4 \pi
\varepsilon_0 r^3}.
\label{Vdd}
\end{eqnarray}
We do not consider the magnetic dipole-dipole interaction since it is of the 
order of $\alpha^2 \approx 10^{-4}$ smaller than the electric
dipole-dipole interaction.
Because the molecules are identical (same isotope, same mass), 
we construct an overall wavefunction $\Psi$ of the system
for which the molecular permutation operator $P$
gives  $P \, \Psi = \epsilon_P \, \Psi$
with $\epsilon_P=+1$
for bosonic molecules and $\epsilon_P=-1$
for fermionic molecules. 
In this study, we consider $^{16}$OH 
bosonic molecules so that $\epsilon_P=+1$. This is due to a total 
spin $\vec{f}=\vec{j}+\vec{i}$ with integer 
quantum numbers $f=1,2$, where $\vec{i}$ is the 
nuclear spin of the molecule ($i=1/2$) and $\vec{j}$ the total angular 
momentum ($j=3/2$) of the molecule considered.
We assume that the nuclear spin $i$ and the angular momentum $j$ 
are decoupled. This is a good approximation for strong
magnetic fields ($\mu B \gg \Delta_\text{hf}$) or high collision energies 
($E_c \gg \Delta_\text{hf}$) where $\Delta_\text{hf} \approx 4$~mK is the 
hyperfine energy splitting between the $f=1$ and $f=2$ manifolds. For the 
state we consider, this assumption is valid for most of the results presented
here especially when $B \gg  100$~G. 
Note that the hyperfine 
structure was previously considered in OH cold collisions that focused on 
smaller collision
energies and
smaller magnetic
fields~\cite{Avdeenkov_PRA_66_052718_2002,Ticknor_PRA_71_022709_2005}.

\begin{figure} [h]
\begin{center}
\includegraphics*[width=6cm,keepaspectratio=true,angle=-90]{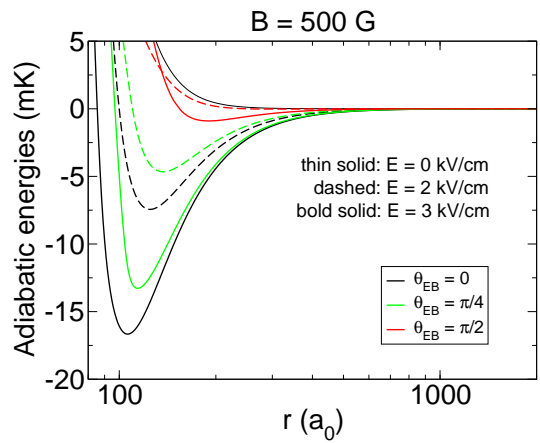}
\caption{(Color online) Adiabatic energies correlating to the $|8\rangle +
|8\rangle$ combined molecular state as a
function of the intermolecular 
separation $r$, for $\theta_{EB}=0$ (black curve), 
$\theta_{EB}=\pi/4$ (green curve), $\theta_{EB}=\pi/2$ (red curve) and 
for different electric fields (thin solid: $E = 0$~kV/cm,
dashed: $E = 2$kV/cm, thick solid: $E = 3$~kV/cm).
The magnetic field is $B=500$~G.
\label{SPAG-FIG}
}
\end{center}
\end{figure}

We construct symmetrized states of the internal wavefunction of the 
combined molecular states $|i_1\rangle \, |i_2\rangle$ of two OH molecules 
with energies $\varepsilon_{i_1} + \varepsilon_{i_2}$ ($i_1,i_2 = 1,...,8$)
\begin{eqnarray}
|i_1, i_2, \eta \rangle = \frac{1}{\sqrt{2(1+\delta_{i_1, i_2})}} 
\bigg[ |i_1, i_2 \rangle + \eta |i_2 \, i_1 \rangle \bigg]
\label{CMSsym}
\end{eqnarray}
for which $P \, |i_1, i_2, \eta \rangle = \eta \, |i_1 , i_2, \eta \rangle$.
$\eta$ is a good quantum number and is conserved during the collision.
If the molecules are in the same molecular internal state, 
only the symmetry $\eta=+1$ has to be considered. If they are in different 
internal state,
both symmetries $\eta=\pm1$ have to be considered. 
As we consider both initial OH molecules in their highest eigenstate 
$|8\rangle$ in the combined electric and magnetic field, the molecules 
are indistinguishable and $\eta = +1$.
The total wavefunction $\Psi(\vec{r})$ with $\vec{r} = \{r,\theta,\varphi\}$ 
is expanded onto a basis set of spherical harmonics $Y_{l,m_l}(\theta,\varphi)$ 
corresponding to the orbital angular momentum of the colliding particles
\begin{eqnarray}
\Psi_{k}(r,\theta,\varphi) &=& \sum_{k''=1}^{N_\text{tot}}  
\frac{1}{r} \,  F_{k'' k}(r) \, Y_{l'',m_l''}(\theta,\varphi) 
\, |i_1'' \, i_2'', \eta'' \rangle \nonumber \\
& = & \sum_{k''=1}^{N_\text{tot}}  \frac{1}{r} \,  F_{k'' k}(r) 
\, |i_1'' \, i_2'', l'', m_l'', \eta'' \rangle.
\label{Psisph}
\end{eqnarray}
where $k=i_1,i_2,l,m_l,\eta$ and $N_\text{tot}$ is the total number 
of diabatic channels we use in our calculation.

Symmetry consideration can restrict the number of channels required.
Because $\eta = +1$, and to satisfy  $P \, \Psi = \epsilon_P \, \Psi$ with 
$\epsilon_P = +1$,  $l$ must take even values. In this study 
we consider only the partial waves $l=0,2,4,6,8$, 
finding these sufficient to converge the results at the collision energies
investigated here. The total number of channels is
typically $N_\text{tot} = 1620$. 
Moreover, when the fields are parallel, the total quantum number 
$M=m_{j_1}+m_{j_2}+m_l$ is conserved, and $N_\text{tot} = 118 $ for the 
$M=+3$ 
components for the initial states $m_{j_1}=3/2$, $m_{j_2}=3/2$, $m_l=0$ for example.
The total energy $E$ is equal to the sum $\varepsilon_{i_1} + \varepsilon_{i_2} + E_c$, 
where $E_c$ is the initial collision energy. 
The total energy
$E$ is conserved during the collision. We choose the zero of energy 
to be equal to $\varepsilon_{8} + \varepsilon_{8}$, the energy of our 
initial molecular states.
The time-independent Schr\"odinger equation $H \Psi = E \Psi$ 
provides a diabatic set of close-coupling differential equations for the 
radial functions $F_{k' k}(r)$ from a state $k$ to a state $k'$
%
%
\begin{multline}
\left\{ - \frac{\hbar^2}{2 m_\text{red}} \frac{d^2}{d r^2} 
 + \frac{\hbar^2 \, l(l+1)}{2 m_\text{red} r^2} - E \right\} 
\, F_{k' k}(r) \\
+ \sum_{k''=1}^{N_\text{tot}} 
  {\cal U}_{k' k''}(r) 
\  F_{k'' k}(r) = 0
\label{eqcoup}
\end{multline}
where
\begin{eqnarray}
{\cal U}_{k' k''}(r) = 
\langle i_1', i_2', l', m_l', \eta' | \, V_\text{dd} \, | i_1'', i_2'', l'', m_l'', \eta'' \rangle. 
\label{Umatrix}
\end{eqnarray}
Using Eq.~\eqref{CMSsym} and the fact that the individual molecular 
eigenstates $|i_1\rangle$ and $|i_2\rangle$ in Eq.\eqref{MSsym} are linear combinations 
of the basis set $|j_1, m_{j_1}, |\omega_{j_1}|, \epsilon \rangle$ 
and $|j_2, m_{j_2}, |\omega_{j_2}|, \epsilon \rangle$ after diagonalisation,
we can obtain the coupling matrix elements ${\cal U}_{k' k''}(r)$
knowing that the dipole-dipole interaction in the Hund's case (a) molecule-molecule 
basis set 
$ |j_1, m_{j_1}, \omega_{j_1}, j_{2}, m_{j_2}, \omega_{j_2}, l, m_l \rangle$  
is expressed by
%
%
\begin{multline}
\langle j_1, m_{j_1}, \omega_{j_1}, j_{2}, m_{j_2}, \omega_{j_2}, l, m_l | \\
 \, V_\text{dd} \,  
| j_1', m_{j_1}', \omega_{j_1}, j_{2}', m_{j_2}', \omega_{j_2}, l', m_l'
\rangle \\
= - \frac{\sqrt{30} \, d_1 d_2}{4 \pi \varepsilon_0 r^3} \,
 (-1)^{m_{j_1}-\omega_{j_1}} \, (-1)^{m_{j_2}-\omega_{j_2}}  \, (-1)^{m_l}  \\
\sqrt{2j_1+1}  \, \sqrt{2j_2+1} \, \sqrt{2l+1}   \\
\sqrt{2j_1'+1} \, \sqrt{2j_2'+1} \, \sqrt{2l'+1}  \\
\sum_{p=-2}^{2} \, \sum_{p_1=-1}^{1} \, \sum_{p_2=-1}^{1} \, \left(
\begin{array}{ccc} 1 & 1 & 2 \\ p_1 & p_2 & -p \end{array} \right) \\
\, \left( \begin{array}{ccc} j_1 & 1 & j_1' \\ -\omega_{j_1} & 0 & \omega_{j_1}
\end{array} \right)
\, \left( \begin{array}{ccc} j_1 & 1 & j_1' \\ -m_{j_1} & m_{j_1} - m_{j_1}' &
m_{j_1}' \end{array} \right) \\
\, \left( \begin{array}{ccc} j_2 & 1 & j_2' \\ -\omega_{j_2} & 0 & \omega_{j_2}
\end{array} \right)
\, \left( \begin{array}{ccc} j_2 & 1 & j_2' \\ -m_{j_2} & m_{j_2} - m_{j_2}' &
m_{j_2}' \end{array} \right) \\
\, \left( \begin{array}{ccc} l & 2 & l' \\ 0 & 0 & 0 \end{array} \right)
\, \left( \begin{array}{ccc} l & 2 & l' \\ -m_{l} & m_{l} - m_{l}' & m_{l}'
\end{array} \right) .
\end{multline}

The multichannel interaction is illustrated in Fig.~\ref{SPAG-FIG} by
showing the lowest adiabatic energies of the symmetrized 
combined molecular state $|8, 8, \eta=+1\rangle$ as a function of $r$ at a 
magnetic field of $B=500$~G, for the electric fields $E=0,2,3$~kV/cm and different 
orientations $\theta_{EB}=0,\pi/4,\pi/2$. These curves are obtained by diagonalizing 
the matrix ${\cal U}(r)$ for each $r$. The thin solid black line shows the
result for zero electric field $E=0$~kV/cm 
illustrating its repulsive $C_6/r^6$ behavior~\cite{Stuhl_N_492_396_2012}.
When the electric field is turned on, second-order perturbations induce 
an attractive $C_4/r^4$~\cite{Avdeenkov_PRA_66_052718_2002,Quemener_PRA_84_062703_2011} coming
from a mixing of the $l=0$ and $l=2$ 
partial waves of the dipole-dipole interaction. 
This interaction  becomes increasingly attractive as the electric field grows,
and also when the magnetic  field is more parallel to the electric field. 
This fact follows qualitatively from the induced electric dipoles of
these state, as discussed for Fig.~\ref{NRG-FIG}: for perpendicular
fields, the electric dipole moment is 
harder to induce hence a weaker attractive $C_4/r^4$ interaction, while for parallel 
fields the reverse is true.

The set of coupled equations Eq.~\eqref{eqcoup} is solved for each $r$
using a diabatic method, using the standard method of the propagation of the 
log-derivative matrix~\cite{Johnson_JCP_13_445_1973}. Matching the log-derivative 
matrix with asymptotic solutions at large $r$ yields finally the scattering matrix 
$S$ and the elastic and total inelastic cross sections and rate coefficients,
which are 
shown in the next section. 
Note that ab initio calculations are not precise enough at such cold
temperatures 
($T \le 1$~K) to rule out the presence of a potential energy barrier in the
entrance channel that could prevent the chemical reaction 
OH +OH $\to$ O + H$_2$O to occur~\cite{Ge_ACP_855_253_2006}.
As a consequence, we do not consider the possibility of chemical reactions in
the present study but could be considered in future works by using an absorbing
condition at short range for
example~\cite{Quemener_PRA_84_062703_2011}.

\section{Application to cold and ultracold OH + OH molecular scattering}

To better orient the discussion of scattering, consider the scales of electric
and magnetic fields as seen by the OH molecule.  For a prototypical cold
collision energy $E=1$mK, this energy is the same as the Stark energy
$|\vec{d} \cdot \vec{E}|$ for
a dipole moment $|\vec{d}|=1.67$~D and an electric field $E = 0.0247$ kV/cm, and it is
the same as the Zeeman energy $|\vec{\mu} \cdot \vec{B}|$ for a dipole moment
$|\vec{\mu}|=2 \, \mu_B$ in a magnetic field $B=7.4$~G. As a rule of thumb, one might therefore expect the
magnetic field to have a dominant effect when $B/E \gg 300$ G/(kV/cm), and the
electric field to have a dominant effect when the reverse is true.

The prospect of manipulating collisions via both electric and magnetic fields,
possibly pointing in different directions, to say nothing of different collision
energies, opens a large parameter space to consider.  In this section we will
explore different slices through this parameter space, by varying separately the
electric field magnitude, the magnetic field magnitude, and the relative
orientation of the fields.

\subsection{Scattering versus electric field}

\begin{figure} [h]
\begin{center}
\includegraphics*[width=5cm,keepaspectratio=true,angle=-90]{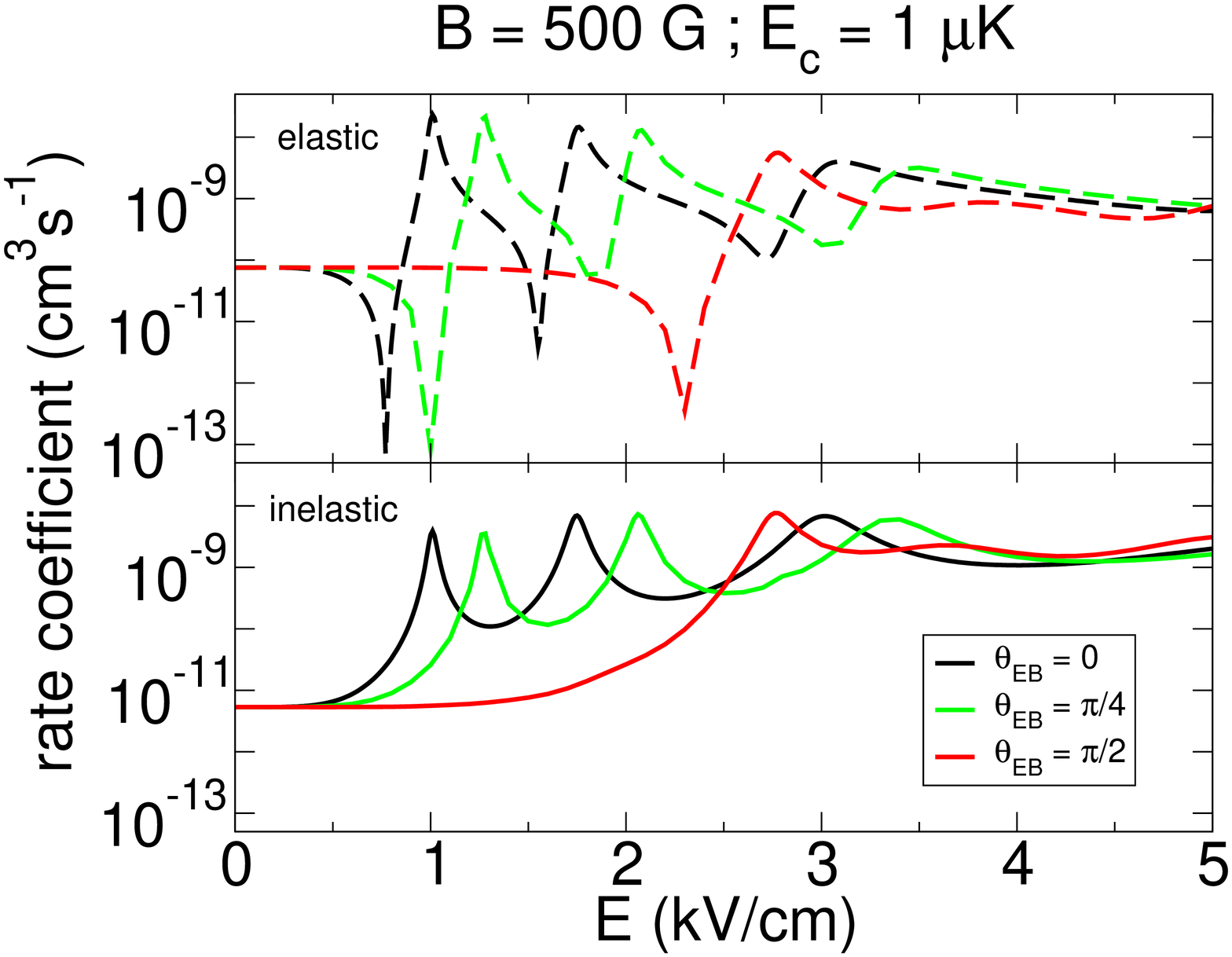}\\
\includegraphics*[width=5cm,keepaspectratio=true,angle=-90]{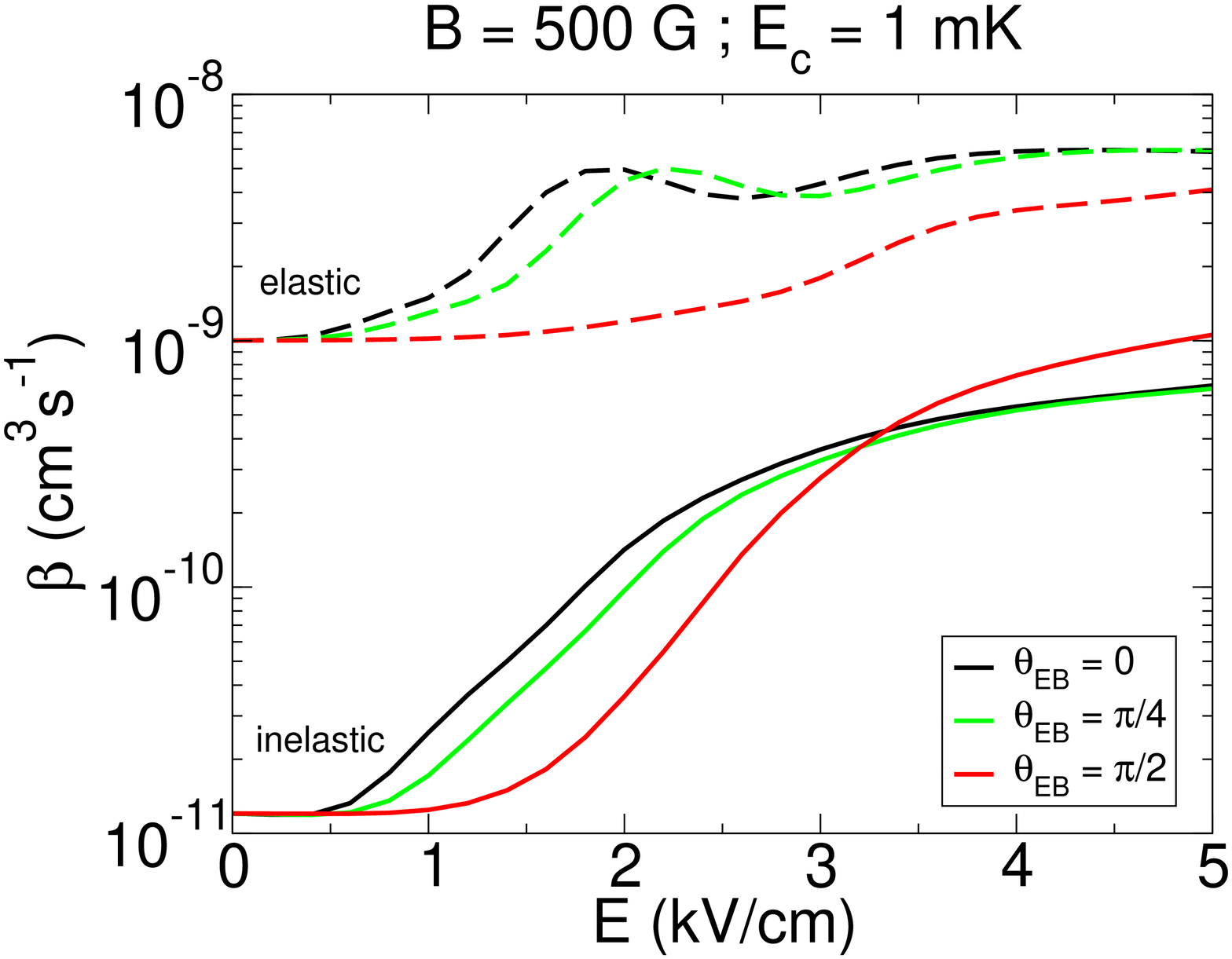}\\
\includegraphics*[width=5cm,keepaspectratio=true,angle=-90]{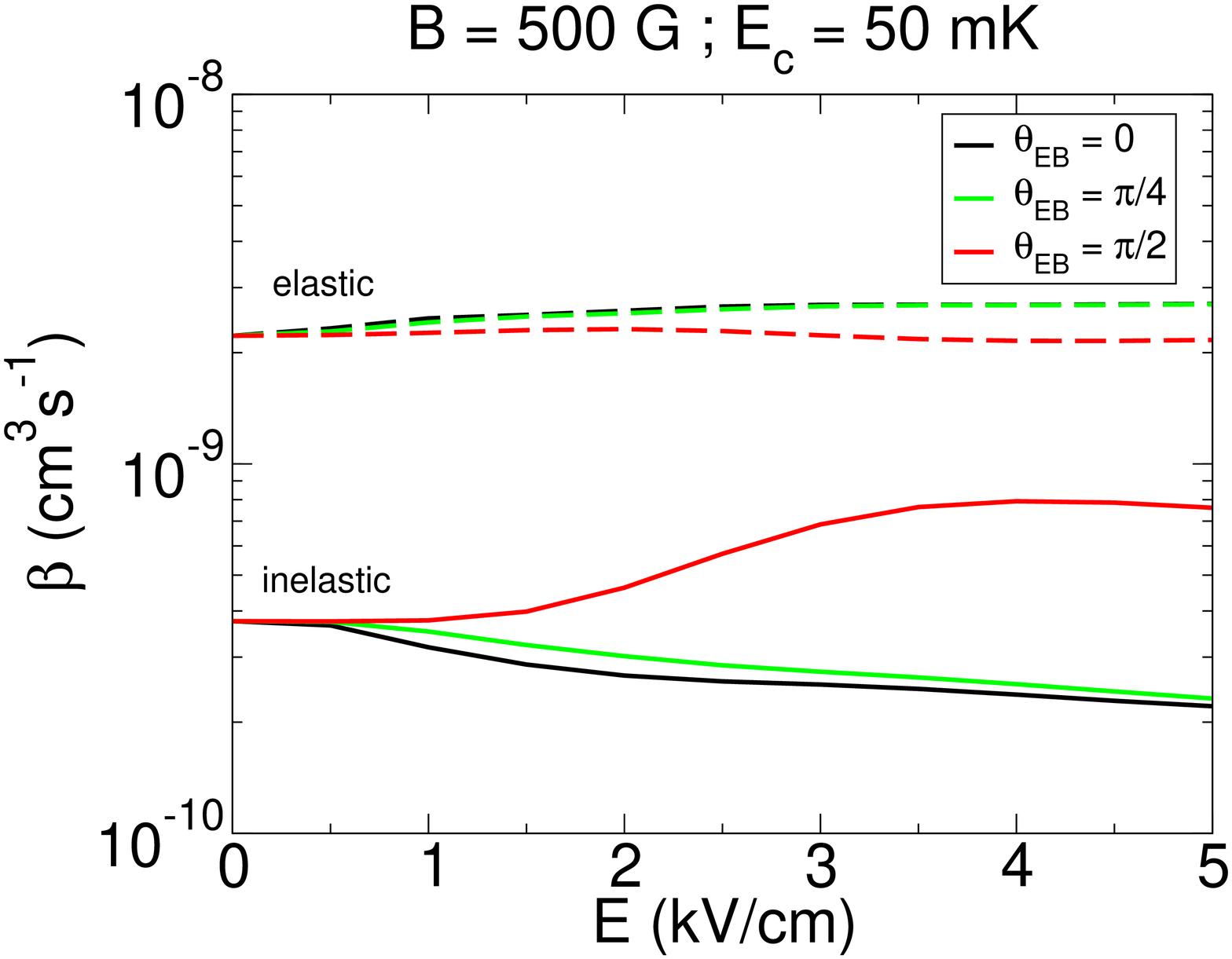}
\caption{(Color online) Rate coefficients as a function of electric field for
$\theta_{EB}=0$ (black curve), 
$\theta_{EB}=\pi/4$ (green curve), $\theta_{EB}=\pi/2$ (red curve). The
collision energy is $E_c=1 \mu$~K (top pannel),
$E_c=1$~mK (middle pannel), $E_c=50$~mK (bottom pannel). The magnetic field is
fixed to $B=500$~G. Elastic process are plotted in dashed lines, inelastic
processes are plotted in solid lines.
\label{RATE-VSEFIELD-FIG}
}
\end{center}
\end{figure}

\begin{figure} [h]
\begin{center}
\includegraphics*[width=5cm,keepaspectratio=true,angle=-90]{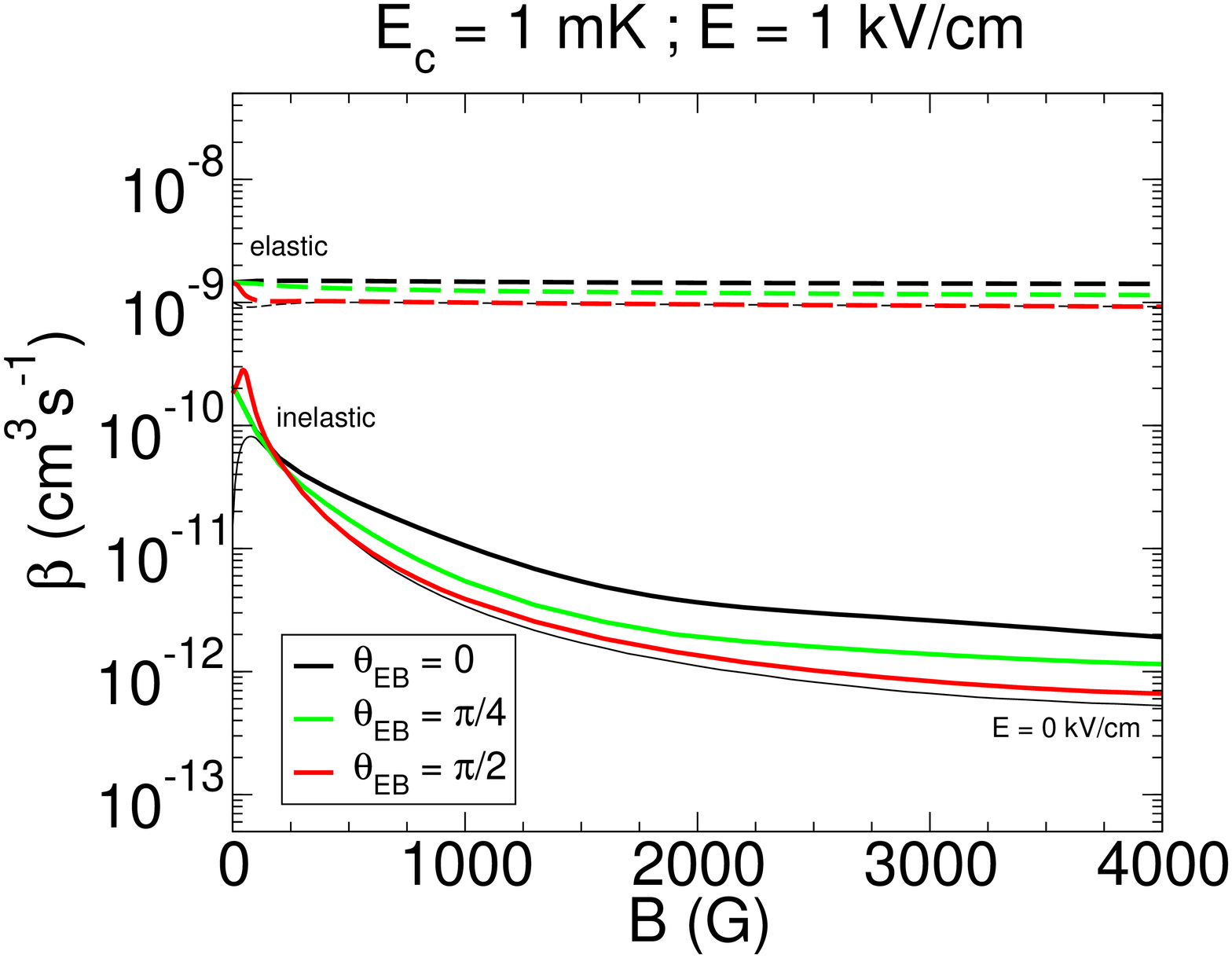}\\
\includegraphics*[width=5cm,keepaspectratio=true,angle=-90]{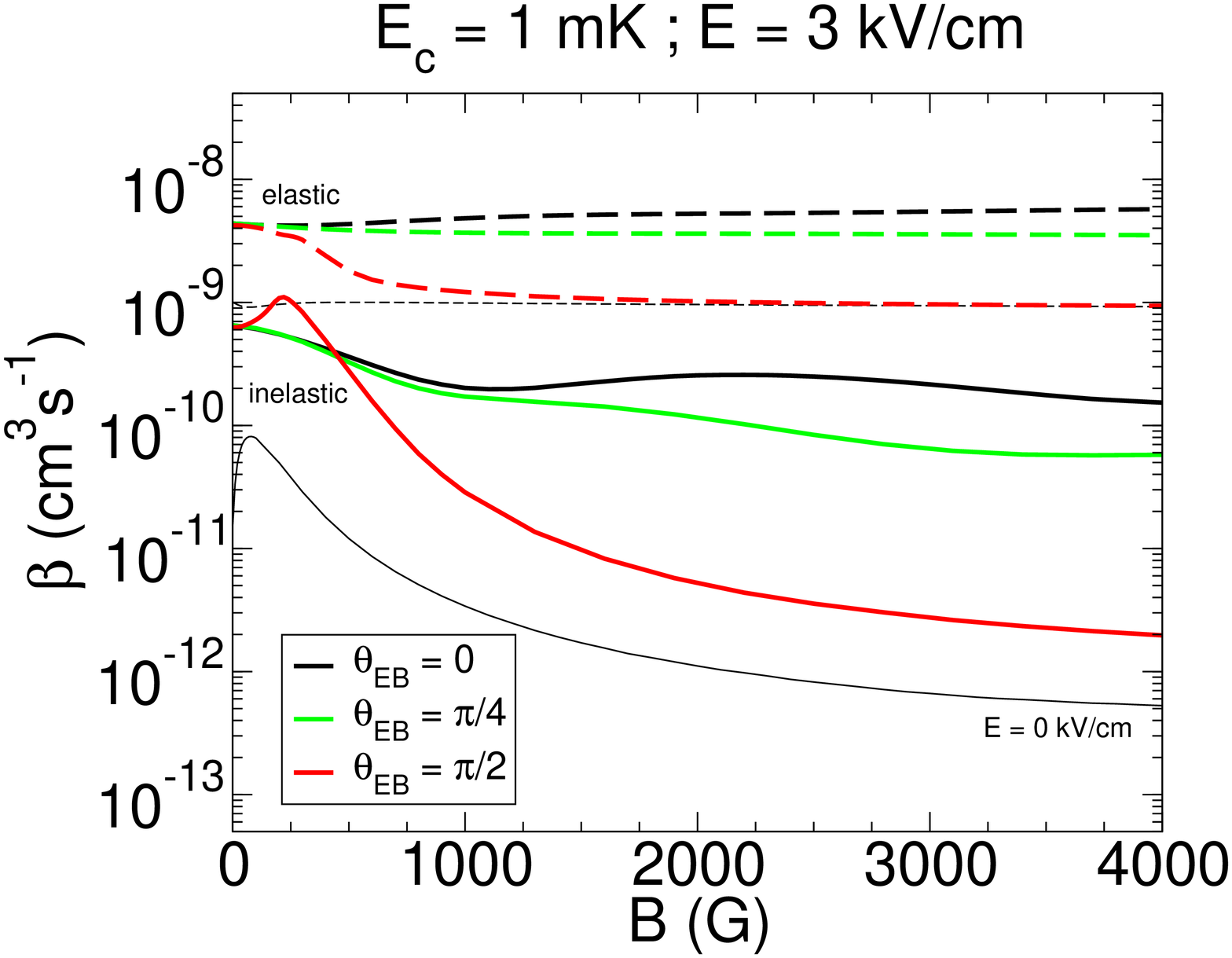}\\
\includegraphics*[width=5cm,keepaspectratio=true,angle=-90]{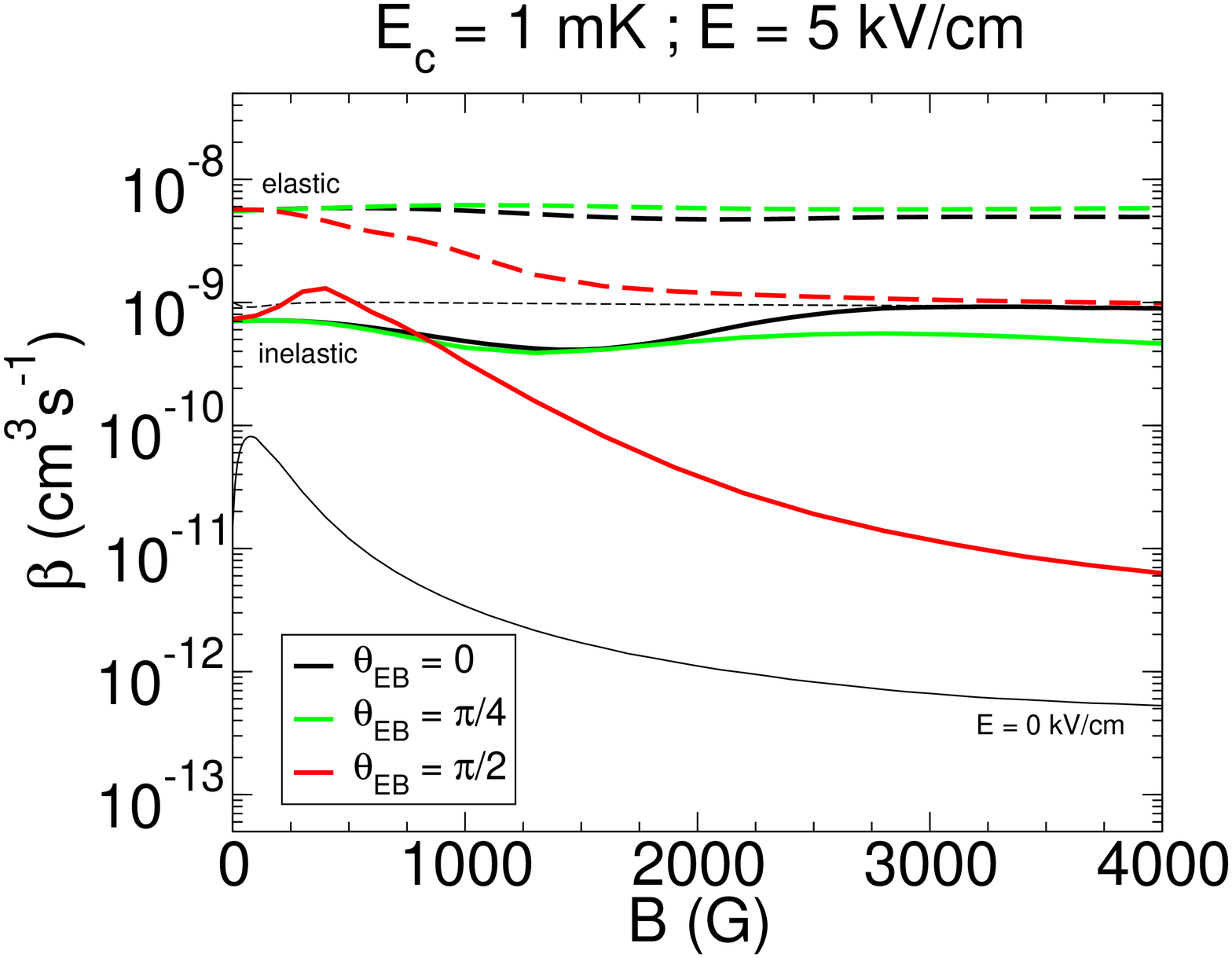}
\caption{(Color online) Rate coefficients as a function of magnetic field for
$\theta_{EB}=0$ (black curve), 
$\theta_{EB}=\pi/4$ (green curve), $\theta_{EB}=\pi/2$ (red curve). The electric
field is $E=1$~kV/cm (top panel),
$E=3$~kV/cm (middle panel), $E=5$~kV/cm (bottom panel). The collision energy
is fixed to $E_c=1$~mK. The thin solid line black line corresponds to $E=0$~kV/cm. 
Elastic process are plotted in dashed lines, inelastic
processes are plotted in solid lines.
\label{RATE-VSBFIELD-FIG}
}
\end{center}
\end{figure}

The general effect of increasing the electric field on scattering of OH is
to increase the inelastic scattering rate.  This effect arises because electric
dipoles are induced, which exert the long-range torques on one another that
drive state-changing collisions~\cite{Avdeenkov_PRA_66_052718_2002}. This effect
is seen in 
Fig.~\ref{RATE-VSEFIELD-FIG} , which
shows the electric field dependence of the elastic 
and inelastic rate coefficients for $\theta_{EB}=0$ (black curves), 
$\theta_{EB}=\pi/4$ (green curves), and $\theta_{EB}=\pi/2$ (red curves) 
at different magnitude of collision energy, for a fixed magnetic field $B = 500$~G. 
The three panels denote three different collision energies.
At an ultra-low collision energy of $E_c=1 \, \mu$~K, the rate coefficients 
show the presence of resonances, signaling the occurrence 
of long-range, ``field-linked'' resonance states predicted in
Ref.~\cite{Avdeenkov_PRA_66_052718_2002}.
They correspond to the coincidence of virtual states 
with the collision energy as the electric field is turned on
and as the adiabatic energy curves become more attractive (seen on Fig.~\ref{SPAG-FIG}).

The electric field values at which these resonances appear clearly depends
on the angle $\theta_{EB}$ between the fields. As noted above, for more
parallel fields it is easier to induce the electric dipole moments 
of the molecules and to obtain strong attractive interaction curves, 
hence resonances appear at  lower electric fields. Vice versa, for more 
perpendicular fields the same resonances appear at higher electric fields 
 since it is harder to induce the electric dipole moments and to obtain 
attractive interaction curves.
Because of these resonances, the inelastic processes can be decreased by three 
orders of magnitude (see for instance at $E = 1$~kV/cm) between the parallel 
and perpendicular cases. The overall trend is a rise of the rates with the 
electric fields due to the increased dipolar coupling with other inelastic 
states as the electric field is turned on.

At a somewhat higher collision energy of $E_c=1$~mK, the resonances are
smoothed out since 
the width of scattering resonances usually increases as the collision energy 
increases~\cite{Mayle_PRA_87_012709_2013}. The rise of
the rates with the 
electric field is still visible. 
According to our rule of thumb, the electric field should exceed $\sim
1.7$~kV/cm to exert a stronger influence on the molecules than a 500~G magnetic
field. And indeed, the second and third panels of
Figure.~\ref{RATE-VSEFIELD-FIG} show suppressed inelastic rates blow about this
field, and enhanced rates above it. Details of the fields still matter, however.
At still larger collisions energy, 50 mK, inelastic rates continue to be
suppressed at high electric field, when the fields are not perpendicular.

\subsection{Scattering versus magnetic field}

As opposed to electric fields, magnetic fields tend to decrease the rate of
inelastic scattering.  This decrease is tied to the general separation of
molecular states in a field, which reduces the Franck-Condon factors between
initial and final states~\cite{Ticknor_PRA_71_022709_2005}.  This effect,
including its modifications due to the electric field, are shown in
Fig.~\ref{RATE-VSBFIELD-FIG}.  
This figure presents the magnetic field dependence of the elastic 
and inelastic
rate coefficients for $\theta_{EB}=0$ (black curves), $\theta_{EB}=\pi/4$ 
(green curves), 
and $\theta_{EB}=\pi/2$ (red curves) for a fixed collision energy of $E_c = 1$~mK. 
In this figure each panel shows the result at a different electric field.
The $E=0$~kV/cm case is represented in thin black line and of course does not depend 
on $\theta_{EB}$. 

The elastic rates are independent of the magnetic field 
while the inelastic rates decrease with the magnetic field, in agreement 
with previous results of Ref.~\cite{Ticknor_PRA_71_022709_2005}. 
Considering the overall trends, our rule of thumb would suggest that 
the magnetic-field suppression would become important at magnetic fields of
300~G, 900~G, and 1500~G, for the three panels, respectively, and this is
approximately what is seen.
For $E=1$~kV/cm, the trends of the rates are comparable and do not differ 
so much from the zero field case. 
The electric field generates the largest deviation from the field-free case
when the fields are parallel.

Once the electric field is larger, the angle between fields plays a
more significant
role. For $E=3$~kV/cm, the rates of the $\theta_{EB}=0$ and $\theta_{EB}=\pi/4$
case have globally increased with the electric field and show a moderate 
magnetic field dependence while the rates of the $\theta_{EB}=\pi/2$ have 
increased but still show a magnetic field dependence similar to lower electric
fields. This is because the electric field is not strong enough for the
perpendicular case to polarize the electric dipole moment and inelastic rates
are still suppressed.
Finally for $E=5$~kV/cm, the $\theta_{EB}=0$ and $\theta_{EB}=\pi/4$ cases 
show a weak magnetic dependence while $\theta_{EB}=\pi/2$ still shows a 
certain dependence. For this case, one can see here that for strong magnetic fields, 
one needs strong electric fields to get a strong electric dipole-dipole interaction 
to increases the inelastic rates.

\subsection{Scattering versus the relative field orientations}

\begin{figure} [h]
\begin{center}
\includegraphics*[width=6cm,keepaspectratio=true,angle=-90]{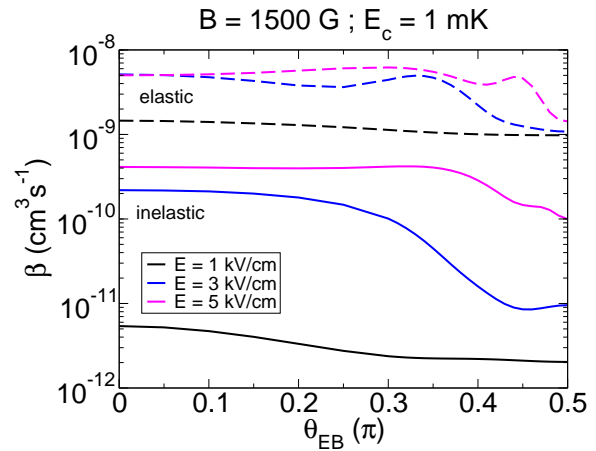}
\caption{(Color online) Rate coefficients as a function of the fields
orientation for $E=1$~kV/cm (black curve), 
$E=3$~kV/cm (blue curve), $E=5$~kV/cm (pink curve). The magnetic field is
$B=1500$~G and the collision energy is $E_c=1$~mK. Elastic process are plotted
in dashed lines, inelastic processes are plotted in solid lines.
\label{RATE-VSTHETAEB-FIG}
}
\end{center}
\end{figure}

Fig.~\ref{RATE-VSTHETAEB-FIG} presents the rate coefficient
for different electric fields at a fixed 
magnetic field $B = 1500$~G and collision energy $E_c=1$~mK,
as a function of the fields angle $\theta_{EB}$.
For an electric field of $E=1$~kV/cm,
elastic and inelastic collisions depend weakly on the angle the electric
field makes with respect to the magnetic field.
The electric field is not strong enough to
play a significant role since $B/E \gg 300$~G/(kV/cm).
Even though, we see that the fields angle can have some effect on the 
inelastic rates, within a factor of 2 to 3.
For an increased electric field of $E=3$~kV/cm, the overall rates have increased 
from the low electric field case
but they decrease from $\theta_{EB}=0$ to $\theta_{EB}=\pi/2$ showing a 
strong anisotropy. Recall that in parallel fields, the magnetic field
helps to polarize the molecules, increasing the inelastic rates.
Molecular collisions highly depend on the relative angle of the fields.
For the same magnetic field, they 
can differ by one order of magnitude for the inelastic processes for example.
Finally at $E=5$~kV/cm, where electric fields have
an effect comparable to magnetic fields ($B/E = 300$~G/(kV/cm)), the inelastic
rates are large and the angle dependence starts to weaken compared 
to the previous case of $E=3$~kV/cm.
In this case the electric field is strong enough
to polarize the molecules along itself, 
so the strength and direction of the magnetic field starts to become irrelevant.

\subsection{Links with experiments}

In experiments, polar molecules of OH are trapped in a magnetic quadrupole 
trap with spatially varying magnetic fields~\cite{Stuhl_N_492_396_2012}, 
and can be simultaneously subject to
a uniform electric field~\cite{Stuhl_MP_2013}. At each
location of the molecules in the trap,
therefore, the molecules experience crossed fields of arbitrary magnitude and
relative orientation, hence different outcomes of two-body collisions 
as previously seen in this study. 
The fact that collision processes highly depend on the position in the trap can
be used to create non-uniform configurations of electric fields in magnetic traps 
so that high value of $\theta_\text{EB}$ regions will favor elastic collisions
while low value regions will favor inelastic ones. With proper field
configurations, 
this could be used as a knife for evaporative cooling in such traps in order to remove 
the particles with the higher energy and keep the particles with the lower energy.

In such traps, it is somewhat more complicated to give an overall rate coefficient
for a given temperature and for a given electric field since the molecular rate 
coefficients depend on the position of the trap as well as the collision energy. 
Experimental data of inelastic loss are now 
available for collision of OH molecules in a quadrupole trap 
and uniform electric fields~\cite{Stuhl_MP_2013}.
To confront these data with theoretical predictions, 
one needs to consider the position and velocity dependent rate coefficients
along with the proper phase space distributions
of molecular positions and velocities to describe molecular losses 
as a function of time.
The option to consider or not the possibility of chemical reactions for OH + OH
collisions will also play a role on the overall magnitude of the loss rates
(inelastic + reactive processes). This has to be kept in mind when comparing
with experimental results.

In addition,
one also needs to consider the time dynamics of the molecules
inside the trap between collisions, since elastic rates 
are higher than the inelastic ones so that
rethermalization plays an important role 
and since elastic scattering 
re-distributes the velocities directions and magnitudes 
of the molecules according to the differential cross sections after a collision.
Such calculations are more complex and usually require Monte-Carlo 
simulations using classical
trajectories~\cite{Barletta_NJP_12_113002_2010,Wu_PRA_56_560_1997}. 
This is beyond the scope of this paper and will be investigated in future
work.

\section{Conclusion}

We have studied the collisions of electric and magnetic polar molecules
in arbitrary configurations of electric and magnetic fields, 
taking the OH molecules as an example. 
The electric dipolar interaction depends on the way the
electric dipole moments are induced. For the state considered in this study,
it is easier to induce the electric dipoles when the electric and magnetic fields
are parallel and this increases the strength of the 
molecule-molecule electric dipolar interaction. 
When the fields become perpendicular, it is harder to induce the electric
dipoles along the electric field axis since the magnetic field tends also to align the 
molecular axis with it. This moderates the strength of the electric dipolar interaction.
This is seen in the dynamics of OH + OH collisions where we found a strong
dependence of the rate coefficients of elastic and inelastic processes 
on the electric and magnetic field configurations. For example, 
more parallel fields increase the inelastic processes while more 
perpendicular fields moderate them.
If the polar molecules are confined in a magnetic quadrupole trap in the presence 
of an electric field, the molecular collisions will
be space and velocity dependent. Therefore the molecular dynamics 
in such traps is complex. Elastic and inelastic collisions 
have to be taken into account as well as the motion of the particles
between collisions in order to describe molecular loss and thermalization.
This will be left for future investigations.

\section*{Acknowledgments}

This material is based upon work supported by the Air
Force Ofﬁce of Scientiﬁc Research under the Multidisciplinary
University Research Initiative Grant No. FA9550-09-1-0588.
This work was also supported in part by the National Science 
Foundation under Grant No. NSF PHY11-25915, during the
``Fundamental Science and Applications of Ultra-cold Polar Molecules" program
held at the Kavli Institute for Theroetical Physics, University of Santa Barbara, USA.
G.Q. acknowledges Triangle de la Physique (contract 2008-007T-QCCM) for financial support.


\end{document}